\documentclass[tighten,nofootinbib,amssymb,floatfix]{article}

\usepackage[DIV=12]{typearea}

\usepackage{t1enc}
\usepackage[footskip=.35in, margin=1in]{geometry}
\usepackage{mathrsfs}
\usepackage{graphicx}
\usepackage{multirow}
\usepackage{hyperref}
\usepackage{bbm}
\usepackage{amsmath}
\usepackage{arydshln}
\usepackage{textcomp}

\usepackage{float}

\usepackage{tikz}
\usetikzlibrary{shadows}
\usetikzlibrary{arrows,shapes,positioning}
\usetikzlibrary{decorations.markings}
\usetikzlibrary{decorations.pathmorphing}
\usetikzlibrary{decorations.pathreplacing}
\tikzstyle arrowstyle=[scale=1]
\tikzstyle directed=[postaction={decorate,decoration={markings, mark=at position .65 with {\arrow[arrowstyle]{stealth}}}}]
\tikzstyle end directed=[postaction={decorate,decoration={markings, mark=at position 1 with {\arrow[arrowstyle]{stealth}}}}]
\tikzstyle reverse directed=[postaction={decorate,decoration={markings, mark=at position .65 with {\arrowreversed[arrowstyle]{stealth};}}}]

\usepackage{authblk}

\usepackage{placeins}
\usepackage{wasysym}

\usepackage[utf8]{inputenc}%for umlaute

\usepackage[small,bf]{caption} %for nice captions
\usepackage{nicefrac} %for nice fractions
\usepackage{subdepth} %for equal height indices

\newcommand{\Dlr}{\mbox{\parbox[b]{0cm}{$D$}\raisebox{1.7ex}
                       {${\,\scriptstyle{\leftrightarrow}}$}}}
		       \newcommand{\MS}{\overline{\mathrm{MS}}}
\newcommand{\Dl}{\buildrel \leftarrow \over D\raise-1pt\hbox{}}
\newcommand{\Dr}{\buildrel \rightarrow \over D\raise-1pt\hbox{}}

\title{
  \flushright
  {\footnotesize DESY-16-223}
  \vspace{5mm}\\
  \centering
  {Gluon momentum fraction of the nucleon from lattice QCD}
  }

\author[a,b]{\normalsize Constantia Alexandrou}
\author[a,b,c]{Martha Constantinou} 
\author[a,b]{Kyriakos Hadjiyiannakou}
\author[d]{Karl Jansen}
\author[a]{Haralambos Panagopoulos}
\author[d]{Christian Wiese}

\footnotesize
\affil[a]{Department of Physics, University of Cyprus, P.O. Box 20537, 1678 Nicosia, Cyprus}
\affil[b]{The Cyprus Institute, 20 Kavafi Street, 2121 Nicosia, Cyprus}
\affil[c]{Temple University, 1925 N. 12th Street, Philadelphia, PA 19122, USA}
\affil[d]{John von Neumann Institute for Computing (NIC), DESY, Platanenallee 6, 15738 Zeuthen, Germany}

\date{\vspace{-15mm}}

\begin{document}
\maketitle

\vspace{15mm}
\begin{abstract}
We perform a direct calculation of the gluon momentum fraction 
of the nucleon, taking into account the mixing with the corresponding
quark contribution. We use maximally
twisted mass fermion ensembles with $N_f=2+1+1$ flavors at a
pion mass of about $370\,\mathrm{MeV}$ and a lattice spacing of $a\approx
0.082\,\mathrm{fm}$ and with $N_f=2$ flavors at the physical
pion mass and a lattice spacing of $a\approx
0.093\,\mathrm{fm}$. We employ stout smearing to obtain a
statistically significant result for the bare matrix
elements.  In addition, we perform a lattice perturbative
calculation including 2 levels of stout smearing to carry
out the mixing and the renormalization of the quark and
gluon operators.  We find, after conversion to the $\MS$
scheme at a scale of $2\,\mathrm{GeV}$,
$\langle x\rangle^R_g {=} 0.284(27)(17)(24)$
for pion mass of about $370\,\mathrm{MeV}$ and 
$\langle x\rangle^R_g {=} 0.267(22)(19)(24)$
for the physical pion mass. In the reported numbers, the first 
parenthesis indicates statistical uncertainties. The numbers in the 
second and third parentheses  correspond to systematic uncertainties 
due to excited states contamination and renormalization, respectively. 
\end{abstract}

\section{Introduction}

The lattice calculation of moments of quark distribution
functions has matured much in the last years, as can
be seen in the reviews of \cite{Alexandrou:2014yha,
Constantinou:2014tga}, for instance. In order to include
disconnected singlet contributions, present works employ
large statistics \cite{Abdel-Rehim:2013wlz,
Alexandrou:2013wca} and even computations for nucleon
observables directly at the physical value of the pion mass \cite{Abdel-Rehim:2015owa}.

For these moments, a complete
non-perturbative renormalization program has been developed
and applied in practice. Furthermore, first attempts to
compute the quark distributions directly on the
lattice have recently been initiated
\cite{Xiong:2013bka,Lin:2014zya,Alexandrou:2015rja}. All
these activities by lattice groups working on nucleon
structure open the exciting prospect that lattice
calculations will eventually provide precise results for
various nucleon moments, charges and form factors with high
statistics and systematic effects under control.

While the computations concerning the quark distribution
functions are approaching a satisfactory situation, the case
of the gluon contributions is much less advanced. In fact,
presently only a few quenched results for the gluon momentum
fraction (GMF) exist\footnote{There has been a recent paper addressing
  the gluon spin contribution in the nucleon
\cite{Yang:2016plb}.}~\cite{Gockeler:1996zg,Horsley:2012pz,Liu:2012nz,Deka:2013zha}. This is a
rather unfortunate situation since the analysis of
phenomenological parton distribution functions data \cite{Alekhin:2013nda} suggests
that at a scale of 6.25\,$\text{GeV}^2$ for instance, all
the quarks only contribute a fraction of about 60 percent to
the total nucleon momentum. This implies that gluons carry
an essential part of the nucleon momentum, in order to
satisfy the sum rule  
\begin{equation}
\sum_q \langle x \rangle_q + \langle x \rangle_g = 1\,. 
\end{equation}
Moreover, the phenomenological estimates of $\langle x
\rangle_g$ have a significantly larger uncertainty than the
corresponding quark moments. The GMF will also be an
important input for the computation of the gluon
contribution to the nucleon spin. 

In this work we perform a calculation of the lowest moment
$\langle x \rangle_g$ of the gluon distribution function
$f_g(x)$ using lattice QCD within the maximally twisted mass
formulation \cite{Frezzotti:2003ni,Frezzotti:2004wz}. We
will use gluon field configurations at a pion mass of about
$370\,\mathrm{MeV}$ but also at the physical pion mass.

The key to obtain results for the GMF is a combination of
high statistics, the use of smeared operators (cf.
\cite{Meyer:2007tm}) and the application of a suitable
renormalization scheme that takes the mixing of the gluon
operator with the corresponding quark singlet operator into
account. The last step is presently done perturbatively but
could be extended non-perturbatively in the future. We will
see that employing these steps will allow us to provide a
quantitative result for $\langle x \rangle_g$ with dynamical
quarks for the first time. A first account of our results
has been discussed in Ref.\;\cite{Alexandrou:2013tfa}.

\section{Theoretical setup}

The gluon momentum fraction of a nucleon state $\langle P \vert$ with 4-momentum
$P^\mu$ can be extracted from matrix elements of the gluonic QCD
energy momentum tensor, see e.g.~\cite{Ji:1994av}
\begin{equation}
\langle P \vert T^{\{\mu\nu\}}_g \vert P \rangle = 2 \langle x \rangle_g P^{\{\mu} P^{\nu\}}\,,
\label{EQN_GM_DECOMP}
\end{equation}
where the normalization $\langle P \vert P \rangle = 2E_N$ 
is used and $\{\dotsm\}$ represents symmetrization and
subtraction of the trace. $E_N$ is the energy of the nucleon.
The gluonic energy momentum tensor itself is defined as
\begin{equation}
T^{\{\mu\nu\}}_g = \frac{1}{4} g^{\mu\nu} G_{\alpha\beta}G^{\alpha\beta} - G^{\mu\sigma}{G^\nu_\sigma},
\end{equation}
where $G_{\mu\nu}=T^a G^a_{\mu\nu}$ is the field strength tensor. 

Based on the conventions used in \cite{Gockeler:1996zg}, 
we construct the gluon operator\footnote{A factor of -2 was added in
order to match the correct decomposition of the Energy-Momentum Tensor.}
\begin{equation}
  \mathcal O_{\mu\nu} = 2\,{\rm Tr}[G_{\mu\sigma} G_{\nu\sigma}]
  \label{EQN_GLUONOP}
\end{equation}
which contains the vector ${\mathcal O_A}_{i}$ 
and scalar ${\mathcal O_B}$ operators  
\begin{equation}
  {\mathcal O_A}_{i}= \mathcal
  O_{i4}\hspace{.5cm}\mathrm{and}\hspace{.5cm}
  \mathcal O_B = \mathcal O_{44} - \frac{1}{3}\mathcal
  O_{jj}\,.
  \label{EQN_GM_OPER}
\end{equation}
Here and in the following equations there is an implicit trace over
the color indices of the field strength tensor and later also the
plaquette term. With Eq.\,(\ref{EQN_GM_DECOMP}) the matrix elements of these
operators can be directly related to the GMF as
\begin{align}
  \label{EQN_intro_2} 
  \langle P \vert \mathcal O_{Ai} \vert P \rangle &= i 4 E_N P_i\langle x \rangle_g\\
  \langle P \vert \mathcal O_B \vert P \rangle  &
= (-4\, E^2_N - \frac{2}{3} {{\mathbf P}}^2)\langle x \rangle_g\,.
  \label{EQN_GMME}
\end{align}
Eq.\,(\ref{EQN_intro_2}) indicates that in
order to extract the GMF from matrix elements of
$\mathcal O_A$, a non-zero momentum for the nucleon fields is
required, whereas the kinematic factor for the operator $\mathcal O_B$
stays finite for zero momentum. 
Thus, for zero momentum the form factor can be extracted as
\begin{equation}
 \frac{\langle P \vert \mathcal O_B \vert P \rangle}{\langle
 P \vert P \rangle}  = - 2\,m_N\, \langle x
\rangle_g\,.
\label{EQN_GM_P0}
\end{equation}

Earlier calculations, see e.g. \cite{Best:1997qp,Guagnelli:2004ga}, 
showed that employing a non-zero momentum
in the definition of the operator corresponding to the 
first moment of the quark distribution leads to a significantly 
enhanced noise-to-signal ratio.  
We therefore have chosen the operator $\mathcal O_B$ for the current calculation. 
We nevertheless plan a test of the operator $\mathcal O_A$
in the future. 

Utilizing Eq.\,(\ref{EQN_GLUONOP}), the operator $\mathcal O_B$ 
can be expressed in terms of the field strength tensor as 
\begin{equation}
  \mathcal O_B = -\frac{4}{3}\left(\sum_{j<k}\,G_{jk}^2  -
\sum_i \,G_{4i}^2\right)\,.
\end{equation}
This expression can now be transferred to the lattice definition 
of the GMF using the operator $\mathcal O_B$ through plaquette terms, 
\begin{align}
  \label{EQN_OP_PLAQ}
  \mathcal O_B=
  -\frac{4}{9}\frac{\beta}{a^4}\left(\sum_i\mathrm{Re}(U_{i4})-\sum_{i<j}\mathrm{Re}(U_{ij})\right)\,.
\end{align}
The operator in Eq.\,(\ref{EQN_OP_PLAQ}) involves two terms which are
very similar in magnitude and have to be subtracted. This points
to the expectation that in order to obtain a precise result a high
statistics and an estimate of the correlation between these two terms
are required.

\section{Lattice calculation}

In \cite{Alexandrou:2013tfa} we discussed  
the approach of employing the Feynman-Hellmann theorem to compute the
gluon momentum fraction. We demonstrated that using the
Feynman-Hellmann theorem is in principle feasible but it would require
a substantial effort to obtain accurate results. 
Thus, we instead follow the path of using the direct computation of the
left-hand side of Eq.\,(\ref{EQN_GM_P0}). This amounts to
computing the ratio of a three- and a two-point correlation
function
\begin{equation}
 R(t,\tau,t') =
  -\frac{1}{2\,m_N}\frac{C^{\text{3pt}}(t,\tau,t'; {\bf P} =
  0)}{C^{\text{2pt}}(t,t'; {\bf P} = 0)} \stackrel{t<\tau<t'}= \langle x \rangle_g \,.
%\frac{2 \beta }{9 a m_N} R\left( N(0);
%\sum_{{\bf x},i<j} \mathrm{Re}(U_{ij}(x)) -
%\sum_{{\bf x},i} \mathrm{Re}(U_{i4}(x)); N(0)\right)\; .
  %\label{EQN_GM_3PT}
  \label{Eq:Ratio}
\end{equation}
The space-time points $({\bf x},t),({\bf x'},t'),({\bf
y,}\tau)$ denote the sink, source and
operator insertion, respectively.

\begin{figure}
  \centering
       \includegraphics[scale=0.9]{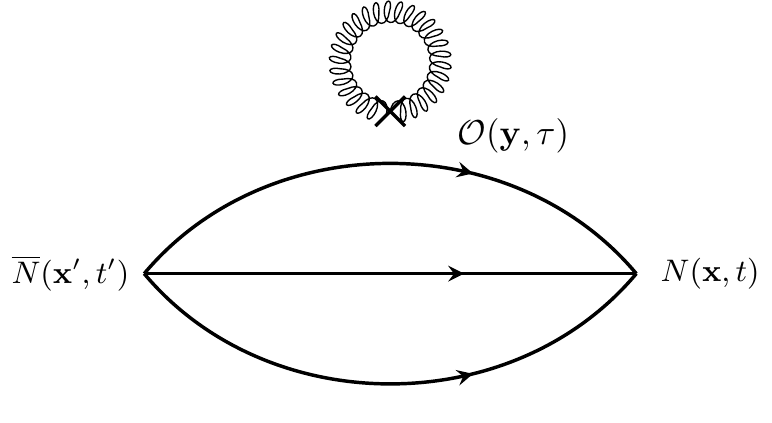}
  \caption{\label{fig_gluon_wick}Schematic picture of Wick contractions
  for the three-point functions with a disconnected gluon
loop.}
\end{figure}

For the GMF, the relevant three-point function is the
expectation value of two nucleon fields and the operator
$\mathcal O_B$ from Eq.\,(\ref{EQN_GM_OPER}), and the two-point
function is defined in the usual way,
\begin{align}
  C^{\text{3pt}}(t,\tau,t'; {\bf P} = 0) &= \sum_{\bf x, y}
\Gamma^+ \left \langle
N(x) \mathcal O_B(y) \overline{N}(x')\right \rangle\,,\\
C^{\text{2pt}}(t,t'; {\bf P} = 0) &= \sum_{\bf x}
\Gamma^+ \left \langle 
N(x) \overline{N}(x')\right \rangle\,,
\end{align}
where $\Gamma^+ = \frac{1+\gamma_4}{2}$ is the parity plus
projector and the standard definition for the nucleon
interpolating fields is used (cf.
\cite{Abdel-Rehim:2015owa}). A schematic picture of the
structure of the three-point function is shown in
Fig.\,{\ref{fig_gluon_wick}}.
\FloatBarrier

Because there are no quark fields in the operator, the
three-point function can be written as the expectation value
of a product of a nucleon two-point function with a gauge
link dependent operator. Generally, we call this a
disconnected correlation function. Consequently, already existing
two-point functions can be re-used while only the gluon
operator has to be calculated on the very same
configurations with a relatively small computational effort.
In order to have an improved signal-to-noise ratio, we
subtract the vacuum expectation value of $\mathcal O_B$ from
the ratio, although strictly speaking this is not necessary
since the expectation value of $\mathcal O_B$ vanishes.

To extract the matrix element of interest three methods
have been employed. The simplest one is the {\it plateau}
method where one must identify a time independent window
in the ratio of Eq.~(\ref{Eq:Ratio}). This method assumes
just one-state dominance. The second method is the
{\it two-state} method, where the first excited state
is taken into account. Inserting a complete set of states
and keeping terms up to the first excited state, the
ratio becomes
\begin{equation}
  R(t,\tau,t') = \frac{A_{00}+A_{01}\left( e^{-\delta E_1(t-\tau)}
    + e^{-\delta E_1(\tau-t')}\right) + A_{11} e^{-\delta E_1(t-t')}}{1+c_1 e^{-\delta E_1(t-t')}},
\end{equation}
where $A_{00}$ is the matrix element of interest and $\delta E_1$ is
the energy gap between the ground state and the first excited state. The
third method, which allows us to control better the excited states,
is called the {\it summation} method. Summing over the insertion time
$\tau$ of the ratio in Eq.~(\ref{Eq:Ratio}), we obtain
\begin{equation}
R^{\rm sum}(t-t') = \sum_{\tau=t'+1}^{(t-1)} R(t,\tau,t') = C + (t-t') A_{00} + \mathcal{O}(e^{-\delta E1 (t-t')})
\end{equation}
where the unphysical contact terms are discarded from the sum. From the slope of the linear fit one can extract
the matrix element.

\section{Lattice setup} \label{SCT_GM}

Our first benchmark calculation is based on 
2298 gluon field configurations 
on a $32^3
\times 64$ lattice from an ETMC (European Twisted Mass
Collaboration) production ensemble \cite{Baron:2010bv},
labeled {\bf B55.32}. It features $N_f=2+1+1$ flavors of
maximally twisted mass fermions, {\it i.e.} two mass degenerate
light quarks and non-degenerate strange and charm quarks.
The ensemble has a bare coupling corresponding to
$\beta=1.95$, which yields a lattice spacing of $a\approx
0.082$\,fm \cite{Alexandrou:2014sha} and the twisted mass
parameter $a\mu = 0.0055$, which corresponds to a pion mass
of $m_{PS} \approx 370$\,MeV. For the two-point function, 15
different source positions are used on each of the 2298
gauge field configurations.  This sums up to 34470
measurements, each for proton, neutron and two different
time directions. 

We also include a second ensemble obtained at the physical value
of the pion mass \cite{Abdel-Rehim:2015pwa}, which is labeled {\bf
cA2.09.48}. Here $N_f=2$ flavors of maximally twisted mass fermions
are employed, together with a clover term with coefficient
$c_{sw}=1.57551$ on a $48^3 \times 96$ lattice. The bare coupling
corresponds to $\beta = 2.1$, which leads to a lattice spacing of
$a\approx 0.093$\,fm, set with the nucleon mass \cite{Abdel-Rehim:2015owa}. 
The twisted mass parameter is set to $a\mu = 0.0009$, which
corresponds, within errors, to a setup with physical pion masses. The
analysis is done on 2094 configurations with 100 different source
positions each, which amounts to a total of 209400 measurements.
\begin{table}
\centering
\begin{tabular}{c c c c c c c c c c }
\hline 
 & $N_f$ & $\beta$ & $\nicefrac{L}{a},\nicefrac{T}{a}$ & $c_{sw}$ & $\kappa$ & $a\mu$ &
 $m_{\text{PS}}$ & $a$ & measurements \\
 & & & & & & & [MeV] & [fm] & \\ 
\hline
B55.32 & 2+1+1 & 1.95 & 32,64 & 0 & 0.161236 & 0.0055 & 370 &
0.082 & 34470 \\
\hline
cA2.09.48 & 2 &2.1 & 48,96 &1.57551 & 0.13729 & 0.0009 & 130 &
0.093 & 209400 \\
\hline
\end{tabular}
\caption{\label{TAB_ENS} Parameters of two different gauge
ensembles that are used in the computation of the GMF. We also give
the number of measurements used for the computation.}
\end{table}
For the quark fields that make up the nucleon interpolating field,
standard smearing methods (Gaussian and Array Processor Experiment (APE) ) were used, which are
known to increase the overlap of the interpolating fields with the
nucleon ground state while decreasing the overlap with excited states
and thus improving the results for nucleon spectroscopy and structure,
{\it cf.} \cite{Alexandrou:2013joa} and references therein.

\section{Bare results and stout smearing}
\label{ResultsStout}

In our first attempt to compute the GMF directly we applied
the gluon operator $\mathcal O_B$ from
Eq.\,(\ref{EQN_OP_PLAQ}) without any additional smearing.
However, in this setup we were not able to detect any signal
despite the large statistics of 34470 measurements on the
B55.32 ensemble, {\it cf.} Table\,\ref{TAB_ENS}, see Fig.\,2
in \cite{Alexandrou:2013tfa}. 

One possible solution to overcome the low signal-to-noise
problem has been suggested in \cite{Meyer:2007tm}, where the
authors propose to use Hypercubic (HYP) smearing \cite{Hasenfratz:2001hp}
for the gauge links in the gluon operator. However, HYP
smearing is a non-analytic procedure; this fact raises some 
conceptual issues, and it also implies that the
perturbative lattice calculation for the desired
renormalization functions would be very cumbersome. 
In the framework of this work we have tested both HYP 
(up to 5 steps) and stout smearing (up to 10 steps). Results with
increased stout smearing are compatible with result produced
with a smaller number of HYP smearing steps. Increasing the number 
of smearing steps may result in contact-term contamination, which 
should be also assessed. Furthermore, the influence of contact terms 
will be reduced by increasing the source-sink separation. To test 
for this effect we take $t_s$ up to 15$a$ and we find that the results 
are compatible with smaller value, e.g. $t_s{=}10a$. Thus, we expect 
that contact-term contamination is small.

Thus, we switch to stout smearing of the gauge links,
as introduced in \cite{Morningstar:2003gk}. This is an
analytic link smearing technique where the gauge links are
smeared according to
\begin{equation}
U_{\mu}^{(n+1)} = \exp\left(iQ_{\mu}^{(n)}\right)
U_{\mu}^{(n)}\,, 
\end{equation}
where $Q_{\mu}$ is a particular linear combination of
perpendicular gauge link staples that are weighted with the
factor\footnote{This parameter is called $\rho$ in the
original work, but in recent works and also here it is 
labeled as $\omega$.} $\omega$, {\it cf.}
\cite{Morningstar:2003gk} for details. Here, we use 
the isotropic four-dimensional scheme and $\omega$
is tuned so that the plaquette reaches a maximal value for 
a given number of smearing steps. 
\begin{figure}[ht!]
       \centering
       \includegraphics[scale=0.57]{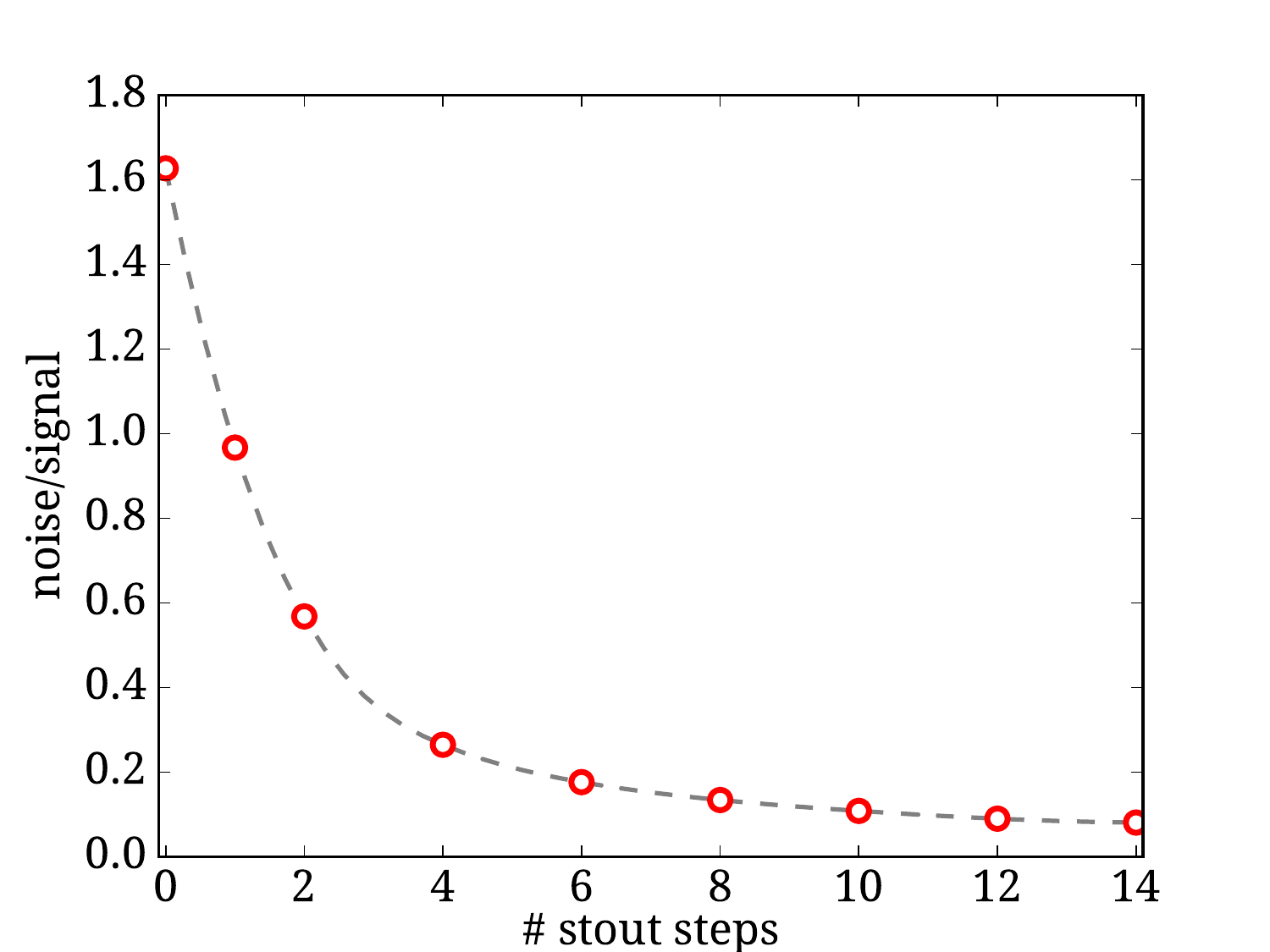}
       \caption{\label{FIG_GM_scaling}Inverse
	 signal-to-noise ratio as a
         function of the number of stout smearing steps. The
         ratio shown here is the average error of
         plateau values divided by the result of a plateau fit for 10
       steps of smearing. All results are given for a source-sink separation of $t_s/a=10$. 
       Here the B55.32 ensemble was used, {\it cf.} Table\,\ref{TAB_ENS}.}
\end{figure}
\FloatBarrier

We tested the effect of stout smearing on the signal-to-noise 
ratio by applying up to 14 smearing steps.  To this
end, we computed the average error of the plateau values for
each level of smearing normalized by the plateau value
that was extracted using 10 steps of smearing. The inverse
signal-to-noise ratio as a function of the number of stout
smearing steps is shown in Fig.\,\ref{FIG_GM_scaling}.

\begin{figure}[h!]
  \centering
  \includegraphics[scale=0.47]{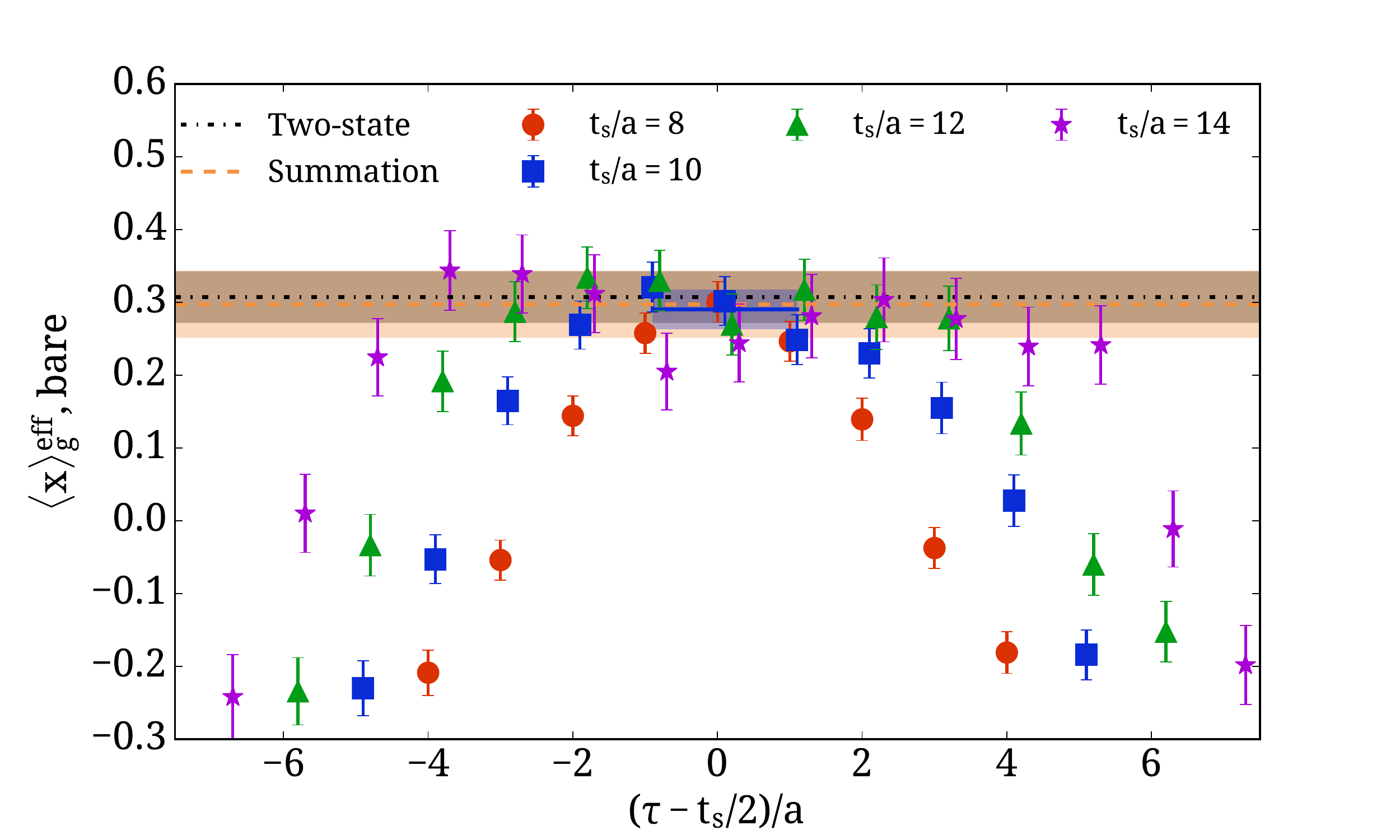}
  \caption{Results for the effective GMF from the B55.32 ensemble as a function of the insertion time-slice $\tau$ for
    four source-sink time separations. Red circles, blue squares, green triangles and magenta stars correspond to
    separations $t_s/a=8,\;10,\;12,\;14$, respectively. The blue band shows the extracted value using
    the plateau method with fit range specified by the band. Results from the two-state (summation) method
  are shown with grey (brown) band spanning the whole x-axis. }
  \label{FIG_GM_B55_ratios}
\end{figure}

\begin{figure}[h!]
  \centering
  \includegraphics[scale=0.47]{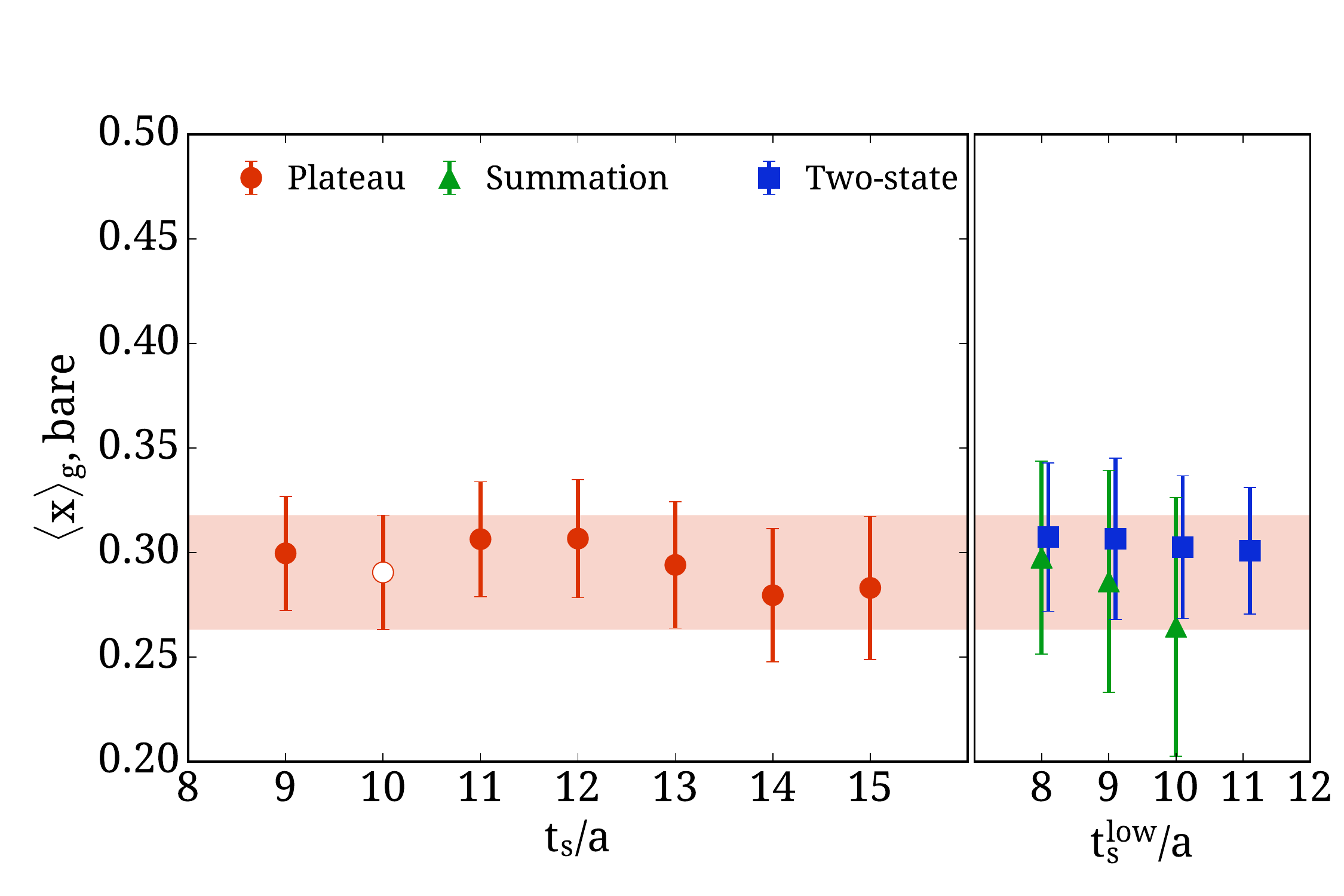}
  \caption{Extracted values for $\langle x \rangle_g^{\text{bare}}$ from B55.32 ensemble using the plateau, two-state and summation methods. The left column shows the extracted values from the plateau method varying the source-sink separation. The open red circle is the value we take as our final value. The right column shows the extracted values using the summation method (green triangles) and two-state fits (blue squares) as one varies the low fit range.}
  \label{FIG_GM_B55_fits}
\end{figure}

From the analysis described above it can be observed that
indeed with an increasing number of stout smearing steps the
signal-to-noise ratio can be substantially improved.  While
the improvement for a smaller number of smearing steps is
quite significant, one notices a saturation for a larger
number of steps. For the B55.32 ensemble, 10 steps of stout
smearing with the parameter $\omega = 0.1315$ are used. The
results for the ratio leading to GMF from this ensemble are shown in
Figs.~\ref{FIG_GM_B55_ratios} - \ref{FIG_GM_B55_fits}.

In order to study the excited state effects we compute
the ratio of Eq.~(\ref{Eq:Ratio}) for various source-sink
time separations. In Fig.~\ref{FIG_GM_B55_ratios}
we present the ratios from where we extract the
matrix element using four separations as one varies
the insertion time-slice using the B55.32 ensemble.
We identify a window where excited states are sufficiently 
suppressed to perform a constant fit using the plateau method 
and we seek for convergence of this value to the ones extracted
using the two-state and summation methods. Our
findings are summarized in Fig.~\ref{FIG_GM_B55_fits}
where several fit ranges are analyzed. We take as our
final value the one for the smallest $t_s$ which
is compatible with the value extracted from the
two-state method. The summation method usually
has larger errors producing results compatible
with the two-state method. Therefore, to
be conservative we provide as a systematic error
due to the excited states the difference between
the plateau value and that extracted from the
two-state fit.

\begin{figure}[h]
  \centering
  \includegraphics[scale=0.47]{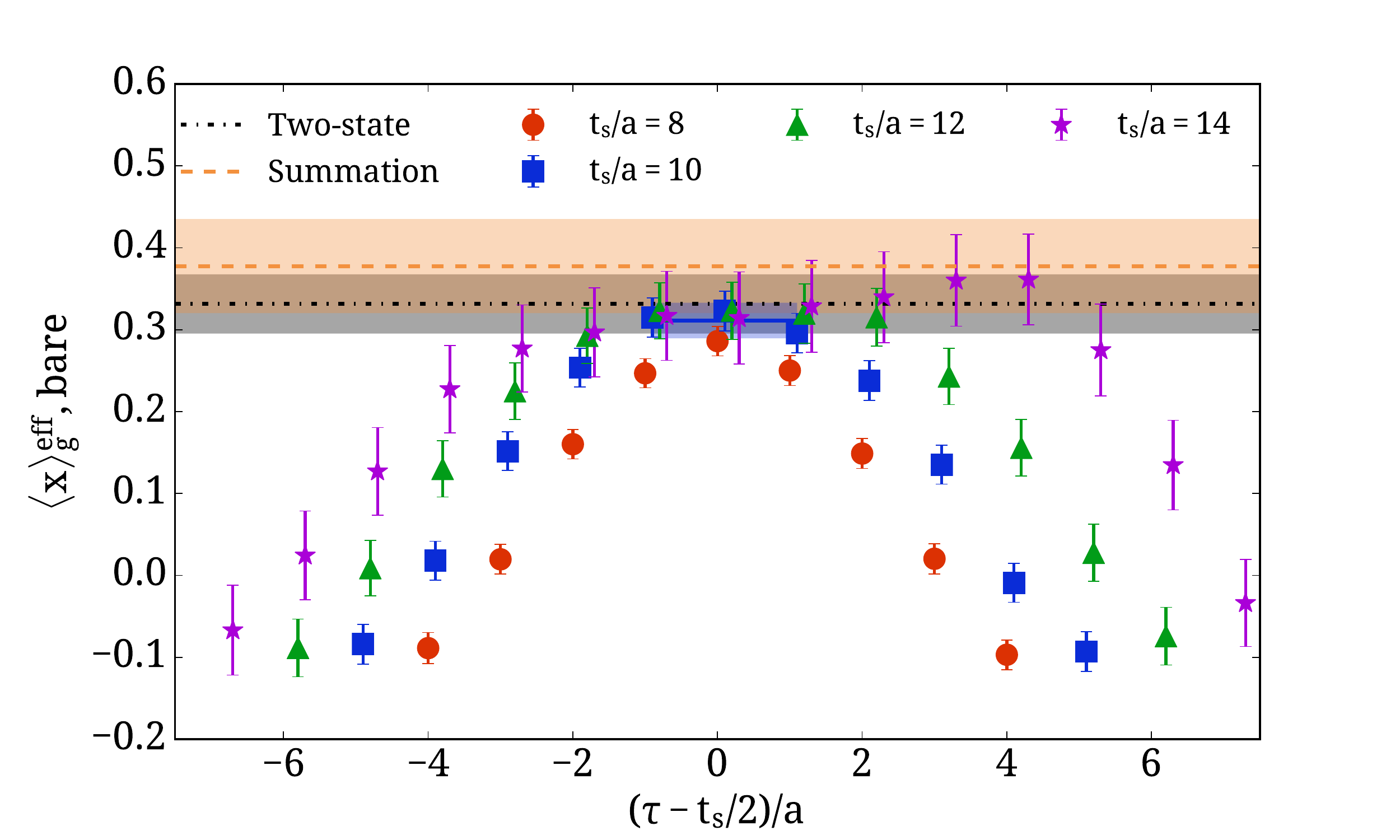}
  \caption{Results for the effective GMF from the cA2.09.48 ensemble. The notation is as in Fig.\ref{FIG_GM_B55_ratios}. }
  \label{FIG_GM_PP_ratios}
\end{figure}

\begin{figure}[h]
  \centering
  \includegraphics[scale=0.47]{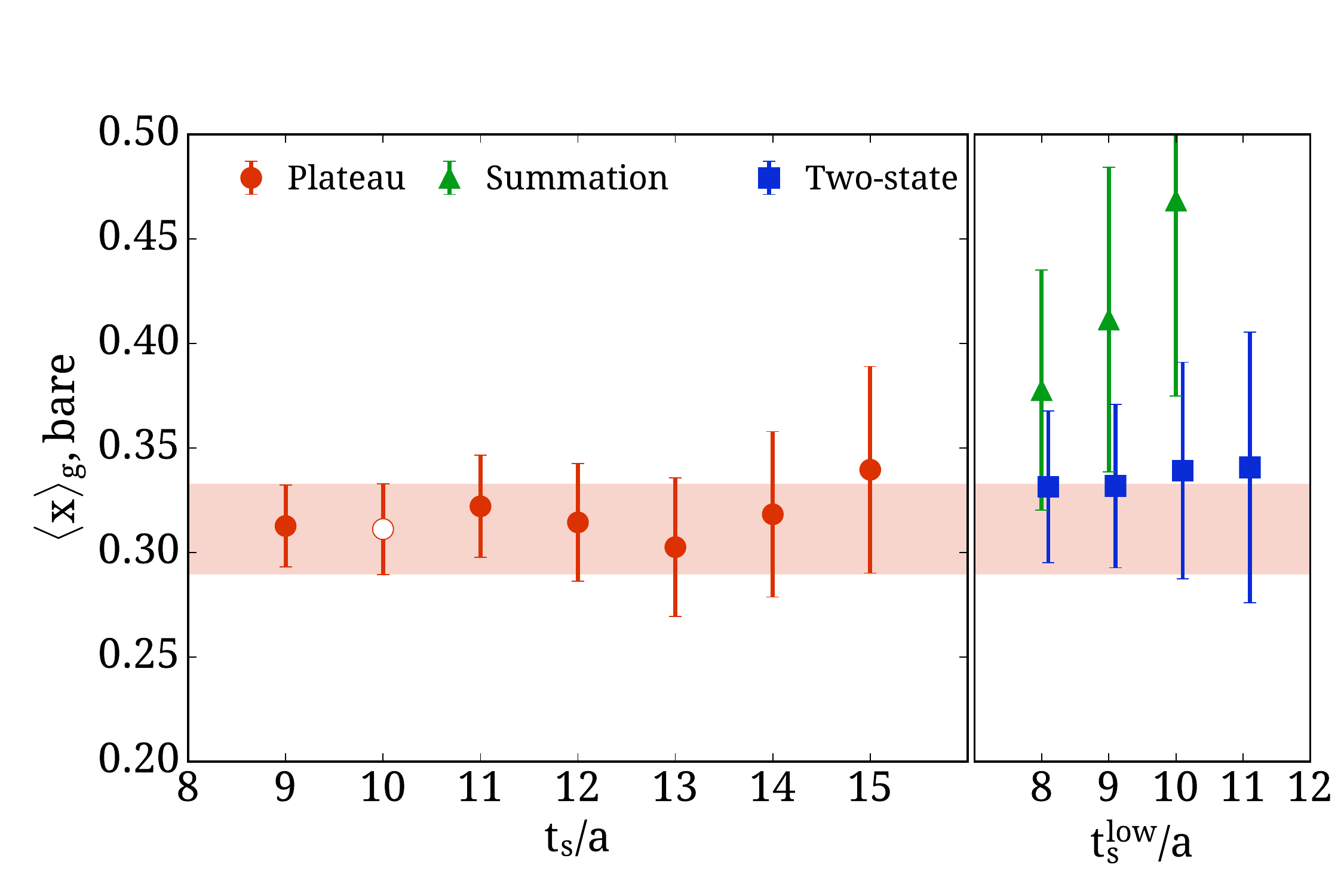}
  \caption{Extracted values for $\langle x \rangle_g^{\text{bare}}$ from the cA2.09.48 ensemble. The notation is as in Fig.\ref{FIG_GM_B55_fits}.}
  \label{FIG_GM_PP_fits}
\end{figure}

%% \begin{figure}
%%        \centering
%%        \includegraphics[scale=0.8]{plot_gm_A09.pdf}
%%        \caption{\label{FIG_GM_A09}{\bf top:} Results for the
%%        effective GMF from cA2.09.48 as a function of the operator insertion time
%%        $\tau/a$ and different source-sink separations
%%        $t_s/a$. 20
%%        steps of stout smearing were applied to the operator.}
%% \end{figure}

The results for the second ensemble with a physical value of the pion mass 
are presented in Figs.~\ref{FIG_GM_PP_ratios} - \ref{FIG_GM_PP_fits}. 
In this case we applied 20 steps of stout smearing with $\omega{=}0.1315$. 
There is no evidence of a large influence of excited states within the statistics 
employed here. For the ensemble at the physical point, we extract the value of
the GMF using the same procedure as the B55.32 ensemble. Our results are as 
follows:
\begin{align}
  {\bf B55.32}    :&\,\,  \langle x  \rangle_{g}^{\text{bare}} = 0.290(27)(17)\,, \nonumber \\
  {\bf cA2.09.48} :&\,\,  \langle x  \rangle_{g}^{\text{bare}} = 0.311(22)(20)\,,
\label{eq_gm_res_bare}
\end{align}
where the number in the first parenthesis is statistical, and the 
second is a systematic due to the excited states contamination.
As mentioned above, the systematic uncertainty is the difference
between the plateau method at $t_s/a{=}10$ and the two-state fit.

\section{Renormalization - Final results}

Yet another challenge regarding the computation of the physical value
of the gluon momentum fraction is the fact that the lattice result has to be
renormalized. Since the gluon operator is a flavor singlet operator,
it will certainly mix with others, the quark singlet operator, for
instance. In total, mixing with operators that are gauge invariant,
Becchi-Rouet-Stora (BRS) variations, or vanish by the gluon equations of motion (e.o.m)
\cite{Joglekar:1976} also appears. Due to this mixing appropriate
renormalization conditions require computation of more than one matrix
element, in order to extract the renormalization factors from a
non-perturbative lattice calculation. This places additional
difficulties compared to the renormalization procedure for other
operators that are relevant for nucleon structure
\cite{Alexandrou:2010me}. Consequently, a different approach has to be
found, and in the framework of this paper we employ a one-loop
perturbative renormalization procedure. In this section we briefly
describe the setup of the calculation and final results needed to
renormalize the GMF. Complete results will appear in a following
publication~\cite{GluonPertRenorm}.

The basis of operators that mix with each other (to one loop) is (see,
e.g., \cite{Caracciolo:1991cp})
\begin{eqnarray}
\label{O1}
{\cal O}_1^{\mu\nu} &=& 2\,{\rm Tr}\left[ G^{\{\mu\rho} G^{\nu\}\rho} \right]  \\
\label{O2}
{\cal O}_2^{\mu\nu} &=& \bar\psi\,\gamma^{\{\mu}\,\Dlr\,\phantom{}^{\nu\}} \psi\\
\label{O3}
{\cal O}_3^{\mu\nu} &=& \frac{1}{\alpha} \Big{[} \left(\partial^\mu A^\nu + \partial^\nu A^\mu \right) \left(\partial^\rho A^\rho \right) -
\frac{1}{2} \delta_{\mu\nu}\left(\partial^\rho A^\rho \right)^2 \Big{]}+ {\rm ghost\,\,terms}\\
\label{O4}
{\cal O}_4^{\mu\nu} &=& \frac{1}{\alpha} \Big{[} - \left(\partial^\mu A^\nu + \partial^\nu A^\mu \right) \left(\partial^\rho A^\rho \right) -
\frac{1}{2} \delta_{\mu\nu} A^\rho \partial^\rho \partial^\sigma
A^\sigma \Big{]}+ {\rm ghost\,\,terms} \\
\label{O5}
{\cal O}_5^{\mu\nu} &=& A^\nu \frac{\delta S}{\delta A^\mu} +  A^\mu \frac{\delta S}{\delta A^\nu}
-\frac{1}{2} \delta_{\mu\nu}  \sum_\rho A^\rho \frac{\delta S}{\delta A^\rho}
\end{eqnarray}
where $\Dlr=(\Dr - \Dl)/2$. ${\cal O}^{\mu\nu}_1$ is the gluon operator 
under study, ${\cal O}^{\mu\nu}_2$ is the corresponding quark operator, 
${\cal O}^{\mu\nu}_3$ and ${\cal O}^{\mu\nu}_4$ are BRS variation
(they only differ by a total derivative) and ${\cal O}^{\mu\nu}_5$
vanishes by the equations of motion. The ghost parts of
operators ${\cal O}^{\mu\nu}_3$ and ${\cal O}^{\mu\nu}_4$ are
irrelevant for this one-loop computation and are not presented here. 
Note that in the calculation we employed traceless operators,
and in such a case there are no lower dimensional two-index traceless
symmetric tensors. Furthermore, we sum over the spatial position of
the operator insertion, resulting in a momentum conservation when
Fourier transforming in the momentum space. The external legs of the
one-loop Feynman diagrams carry the same momentum.

\begin{figure}
       \centering
       \hspace{.5cm}\includegraphics[scale=.66]{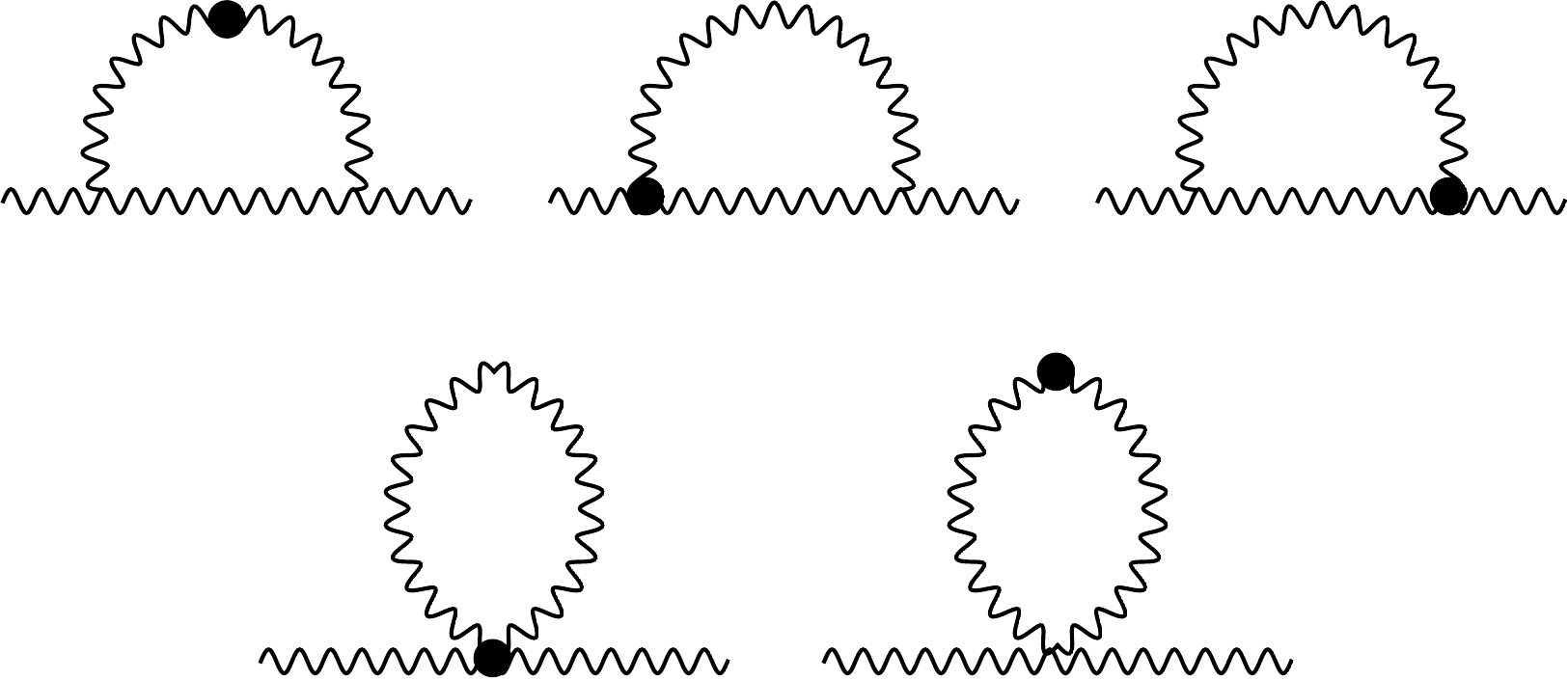}
       \caption{\label{diags_Zgg} One-loop Feynman diagrams contributing
         to the multiplicative renormalization of ${\cal O}_1$.}
\end{figure}

\begin{figure}
       \centering
       \hspace{.5cm}\includegraphics[scale=.45]{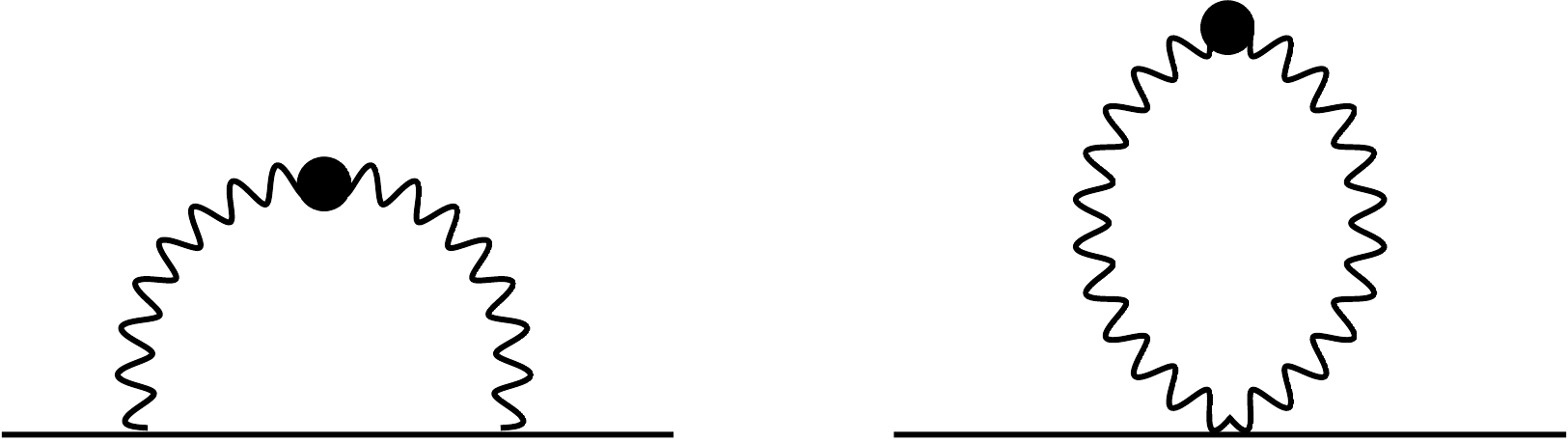}
       \caption{\label{diags_Zgq} One-loop Feynman diagrams contributing
         to the mixing coefficient in ${\cal O}_1$ due to ${\cal O}_2$.}
\end{figure}

From this point forward we concentrate on the singlet case, $\mu=\nu$,
and we drop the Lorentz indices, that is, ${\cal O}_i\equiv {\cal O}_i^{\mu\mu}$ 
($i=1,\cdots,5$). Furthermore, we indicate by ${\cal O}_1$ the combination resulting 
${\cal O}_B$, in order to have the correct mixing coefficients. To identify and extract 
the multiplicative renormalization function of the gluon operator ${\cal O}_1$, one
must construct a mixing matrix with elements that are appropriate Green's
functions of the above operators. However, mixing with ${\cal O}_3$ - 
${\cal O}_4$ vanishes at the one-loop level and the matrix elements of the
operator ${\cal O}_5$ between physical states vanish; the mixing matrix 
simplifies considerably. In particular, the only Feynman diagrams that enter 
our one-loop calculation are those of the operators ${\cal O}_1$ and ${\cal O}_2$, 
within external quarks and gluons. As we are interested in the renormalization of the
operator ${\cal O}_1$ only, we present the relevant Feynman diagrams in
Figs.\,\ref{diags_Zgg}\,-\,\ref{diags_Zgq}.
%\FloatBarrier

\vspace{0.25cm}
%\FloatBarrier
%

The most important consequence of the vanishing physical matrix elements of
${\cal O}_3$ - ${\cal O}_5$ is that the ratio shown in Eq.\,(\ref{EQN_GM_P0})
is a linear combination of contributions from only ${\cal O}_1$ and ${\cal O}_2$. 
Note, however, that to correctly identify the multiplicative renormalization 
of ${\cal O}_1$, the operators ${\cal O}_3$ - ${\cal O}_5$
must be taken into account in the perturbative renormalization
procedure (see Eq.\,(\ref{LggDR})).

To make contact with phenomenological and experimental data, one needs
the renormalization functions in the $\MS$ scheme. An ideal method 
to extract the $\MS$ results is to perform the computation in both 
dimensional (DR) and lattice (L) regularizations; one then extracts all 
relevant renormalization functions by demanding that renormalized lattice 
Green functions coincide with the corresponding ones in (DR), in the 
$a \to 0$ limit (cf. \cite{Constantinou:2015ela} for a similar application). 
Thus, one avoids intermediate schemes. Let us briefly outline this procedure below.

In cases of operator mixing, renormalized operators are related to the
bare ones via $\hat{\cal O}^R=\hat{\cal Z}\, \hat{\cal O}$. In
our case $\hat{\cal Z}$ is a $5\times5$ mixing matrix of the form
\begin{equation}
\hat{\cal Z} = \hat{1} + {\cal O}(g^2)\,,
\end{equation} 
where $g$ is the renormalized coupling constant. In this paper we are
interested in the renormalization of the gluon operator, ${\cal O}_1$,
and we only need to compute the first row of the mixing matrix to
one-loop, which has only two non-zero matrix elements, that is $Z_{11}$
and $Z_{12}$. Alternatively, we write
\begin{equation}
{\cal O}^R_\alpha = \sum_\beta Z_{\alpha\beta} {\cal O}_\beta\,\qquad \alpha,\beta=1,2\,.
\end{equation}
In a more convenient notation, the $X$-$X$ bare amputated Green's functions ($X=1(2)$:
corresponds to a gluon(fermion) field) can be expressed in terms of
the renormalized Green's functions, that is,
\begin{equation}
\langle X {\cal O}_\alpha X \rangle = Z_X^{-1} \sum_\beta \left(Z^{-1}\right)_{\alpha\beta} 
\langle X {\cal O}_\beta X \rangle_R
\end{equation}
where $Z_X$ is the renormalization function of the fermion/gluon
field, defined via
\begin{equation}
\Psi = \sqrt{Z_q} \Psi^R \qquad
A_\nu = \sqrt{Z_A} A_\nu^R 
\end{equation}

\bigskip
\noindent
\underline{\it{Dimensional Regularization}}

\smallskip
Next, we present the results in Dimensional Regularization for the
amputated Green's functions entering the renormalization of the gluon operator,
${\cal O}_1$. The renormalization functions in the $\MS$ scheme in
DR are defined such as to cancel the divergent parts of the matrix
elements. The expressions related to the one-loop renormalization of
the gluon operator reduce to
\begin{eqnarray}
  \Lambda_{11}^{\mathrm{1-loop}} \Bigg{|}_{1/\epsilon} &=& \left(-z_A
-z_{11}\right) \Lambda_{11}^{\mathrm{tree}} 
- z_{31} \Lambda_{31}^{\mathrm{tree}} - z_{41}
\Lambda_{41}^{\mathrm{tree}}
- z_{51} \Lambda_{51}^{\mathrm{tree}} \\
\Lambda_{12}^{\mathrm{1-loop}}\Bigg{|}_{1/\epsilon} &=& -z_{12}
\Lambda_{12}^{\mathrm{tree}} 
%\Lambda_{21}^{1-loop} &=& -z_{21} \Lambda_{21}^{tree} \\
%\Lambda_{22}^{1-loop} &=& \left(-z_{22}-z_\psi\right) \Lambda_{22}^{tree}
\end{eqnarray}
where $\Lambda_{aX}\equiv \langle X {\cal O}_\alpha X \rangle$ and
$z$'s are the one-loop contributions of the corresponding
renormalization functions, that is
\begin{eqnarray}
Z_A = 1 + z_A + {\cal O}(g^4) \\
Z_{ii} = 1 + z_{ii} + {\cal O}(g^4) \\
Z_{ij} = 0 + z_{ij} + {\cal O}(g^4) 
\end{eqnarray}
It should be noted that, modulo a total derivative, the gluon parts of ${\cal O}_3$
and ${\cal O}_4$ coincide ($\Lambda^{\rm tree}_{31} = \Lambda^{\rm tree}_{41}$) 
and, thus, we cannot disentangle $z_{31}$ and $z_{41}$ from the
Green's functions we study. However, this does not affect the extraction of $z_{11}$. 

In our one-loop calculation we find:
\begin{eqnarray}
\label{LggDR}
\Lambda_{11}^{\mathrm{1-loop,DR}} \Bigg{|}_{1/\epsilon} &=&
\frac{g^2}{16\,\pi^2} \frac{N_c}{\epsilon}
\Bigg{[}\Lambda_{11}^{\mathrm{tree,DR}} \left(-\frac{5}{3}
  -\frac{\beta}{2} \right)\nonumber\\
  && \hspace{1cm} - \left(
\Lambda_{31}^{\mathrm{tree,DR}} +
\Lambda_{41}^{\mathrm{tree,DR}} \right) -
2\,\Lambda_{51}^{\mathrm{tree,DR}}\Bigg{]}  \\[2ex]
\Lambda_{12}^{\mathrm{1-loop,DR}} \Bigg{|}_{1/\epsilon} &=&
\frac{g^2}{16\,\pi^2} \frac{N_c^2-1}{\epsilon\,N_c} 
\, \Lambda_{22}^{\mathrm{tree,DR}} \left(\frac{5}{3} +\beta \right) 
\end{eqnarray}

By definition, the finite terms of
$\Lambda_{ij}^{\mathrm{1-loop,DR}}$ do not
appear in the evaluation of $Z_{ij}^{\mathrm{1-loop,DR}}$, but they are key
elements in obtaining $Z_{ij}^{\mathrm{L,\MS}}$ as explained below. 

Let us slightly modify our notation and use the gluon and quark
momentum fraction of the nucleon, $\langle x \rangle_g$ and 
$\langle x \rangle_q$, which are more relevant for this paper. For
demonstration purposes we will represent the mixing of physical matrix
elements as a $2\times2$ matrix
\begin{equation}
  \binom{\langle x \rangle_g}{\sum_q\langle x \rangle_q} =
  \begin{pmatrix} Z_{11} & Z_{12} \\ Z_{21} & Z_{22} \end{pmatrix}
  \binom{\langle x \rangle_g^{\text{bare}}}{\sum_q\langle x
  \rangle_q^{\text{bare}}}\,.
\end{equation}
Thus, the physical result of the gluon momentum fraction can be
related to the non-perturbative results for $\langle x \rangle_g$ and 
$\langle x \rangle_q$ by
\begin{equation}
    \langle x \rangle_g^R = Z_{11} \langle x \rangle_g + Z_{12}\sum_q \langle x \rangle_q\,,
\label{RenormGluon}
\end{equation}
where a certain scheme, {\it e.g.}
$\MS$, and an energy scale $\mu$ have to
be chosen. The expressions for $Z_{11}$ and $Z_{12}$ in DR and in
the $\MS$ scheme are
\begin{eqnarray}
Z_{11} = 1 + \frac{g^2\,N_f}{16\,\pi^2} \frac{2}{3\,\epsilon}\\
Z_{12} = 0 - \frac{g^2\,C_f}{16\,\pi^2} \frac{8}{3\,\epsilon}
\end{eqnarray}
where $C_f=\frac{N_c^2-1}{2N_c}$.

\bigskip
\noindent
\underline{\it{Lattice Regularization}}

\smallskip
To obtain the corresponding lattice results for $Z_{ij}$ in the $\MS$
scheme we will make use of the DR results, so that an indermediate
Regularization independend (RI) type prescription is avoided. Renormalizability of the theory
implies that the difference between the one-loop renormalized and bare
Green's functions is polynomial in the external momentum (of degree 0,
in our case, since no lower-dimensional operators mix); this results in
an appropriate definition of the momentum-independent renormalization
functions $Z_{ij}^{\mathrm{L,\MS}}$. More precisely, for the operators under
study we find to one loop 
\begin{eqnarray}
  \langle A_\nu {\cal O}_1 A_\nu \rangle^{\mathrm{DR,\MS}} - 
  \langle A_\nu {\cal O}_1 A_\nu \rangle^{\mathrm{L}} &=& 
  \left(z_A^{\mathrm{L,\MS}} + z_{11}^{\mathrm{L,\MS}}
\right)\,\Lambda^{\mathrm{tree}}_{11} \nonumber\\ 
&+& \left(z_{31}^{\mathrm{L,\MS}}+z_{41}^{\mathrm{L,\MS}}
\right)\,\Lambda^{\mathrm{tree}}_{31} +
z_{51}^{\mathrm{L,\MS}}\,\Lambda^{\mathrm{tree}}_{51} \,\,\,\,\, \\
\langle \Psi {\cal O}_1 \Psi \rangle^{\mathrm{DR,\MS}} -
\langle \Psi {\cal O}_1 \Psi \rangle^{\mathrm{L}} &=&
z_{12}^{\mathrm{L,\MS}}\,\Lambda^{\mathrm{tree}}_{22}
\end{eqnarray}

It should be noted that the smearing of the operator modifies its
renormalization factor, and thus for a proper renormalization it is
required to apply the same smearing in the perturbative calculation. 
The main technical difficulty in such a case is that the smearing
leads to extremely lengthy expressions for the operator's vertices. For
example, the 4-gluon vertex for two smearing steps with general
smearing parameters, $\omega_1$ and $\omega_2$, contains approximately
335,000 terms. This places severe limitations on the number of
smearing iterations we can apply to the operator. In our computation we 
extract the vertices with up to two stout smearing steps with distinct
parameters. This allows us to compare values of the renormalization
functions for the single- and double-smeared operator. We find that 
increasing the number of smearing steps has small effect on the renormalization
functions. This is due to a combination of the small value of the smearing
parameter and the polynomial dependence on $\omega_1$ and
$\omega_2$. We also note that the perturbative calculation is performed for 
general action parameters, so that the results are applicable for a variety of 
gluon/fermion actions.

The general expressions for $Z_{11}$ and $Z_{12}$ are complicated
$4^{\rm th}$-degree polynomials of $\omega_1$ and $\omega_2$, and
cannot be presented here. Thus, we write them in a compact form, as a
function of the quantities $e^{(i)}_{11 / 12} \equiv e^{(i)}_{11 /
12}(\omega_1,\omega_2)$, which also depend on the gluon action 
parameters
\begin{eqnarray}
  Z^{\mathrm{L,\MS}}_{11} &=& 1 + \frac{g^2}{16\pi^2} \left(\frac{e^{(1)}_{11}}{N_c} + 
e^{(2)}_{11}\,N_f -\frac{2\,N_f}{3}\log(a^2\bar\mu^2) \right)  \\
Z^{\mathrm{L,\MS}}_{12} &=& 0 + \frac{g^2\,C_f}{16\pi^2} \left(e^{(1)}_{12} + 
e^{(2)}_{12}\,c_{\rm SW} +\frac{8}{3}\log(a^2\bar\mu^2) \right) \,.
\end{eqnarray}
The computation of the quantities $e^{(i)}_{11 / 12}$ is the most
laborious part of the perturbative work and required the equivalent of
approximately 40 years of computation on a single CPU. This includes,
among other parts, the integration of the internal loop momentum for
several lattice sizes and the extrapolation to the infinite volume limit.
The numerical results for the multiplicative renormalization function,
$Z_{11}^{\MS}$ and the mixing coefficient, $Z_{12}^{\MS}$, are given in
Table\,\ref{TAB_Z} in the $\MS$ scheme at a scale of $2\,\mathrm{GeV}$. The
statistical errors associated with the infinite volume extrapolation
are smaller than the accuracy presented in the table. One can 
observe that the effect of additional smearing steps tends to become
suppressed. This is due to the polynomial dependence on $\omega_1$ and
$\omega_2$, combined with the fact that their numerical value is very
small. It is expected that the effect of further smearing steps
will be smaller than the difference between the 1- and 2-stout results
shown in Table\,\ref{TAB_Z}. Thus, we employ the renormalization factors
using the 2-stout results to renormalize the matrix element presented
in Section~\ref{ResultsStout}.

\begin{table}
\centering
\begin{tabular}{ c c c c c c c }
\hline \\[-2ex]
\phantom{----} & & $Z_{11}^{\mathrm{L,\MS}}$ &
\,\,\quad\hspace*{0.78cm}\vline & & $Z_{12}^{\mathrm{L,\MS}}$ & \\ \hline
\phantom{----} & 0-stout & 1-stout & 2-stout \hspace*{0.025cm}\vline & 
0-stout & 1-stout & 2-stout  \\
\hline
B55.32      &0.9481  &1.0043  &1.0134 \,\,\vline   &0.1720  &0.0278 &-0.0168 \\
\hline
cA2.09.48   &0.8985  &0.9506  &0.9590  \,\,\vline  &0.1120  &-0.0070  &-0.0436  \\
\hline
\end{tabular}
\caption{Multiplicative renormalization and mixing coefficient for the
gluon operator. Results are given in the $\MS$ scheme at a scale of 
$2\,\mathrm{GeV}$.} 
\label{TAB_Z}
\end{table}

According to Eq.\,(\ref{RenormGluon}) the bare quark momentum fraction
enters the renormalization prescription of the gluon momentum fraction.
The quark contributions have been computed for both the connected and
disconnected diagrams for {\bf B55.32}~\cite{Alexandrou:2013joa,Abdel-Rehim:2013wlz}
and {\bf cA2.09.48}~\cite{Abdel-Rehim:2015owa,Alexandrou:2016tuo,Abdel-Rehim:2016pjw}.
Using the bare results 
\begin{align}
{\bf B55.32}    :&\,\,  \langle x\rangle _{u+d} = 0.603(79) \nonumber \\
{\bf cA2.09.48} :&\,\,  \langle x\rangle _{u+d+s} = 0.722(96)\,,
\label{eq:quarkmoment}
\end{align}
we find the following values for the renormalized gluon momentum
fraction in the $\MS$ at $\mu=2\,\mathrm{GeV}$:
\begin{align}
{\rm{\bf B55.32}}    :&\,\, \langle x\rangle^R_g = 0.284(27)(17)(24) \nonumber \\    
{\rm{\bf cA2.09.48}} :&\,\, \langle x\rangle^R_g = 0.267(22)(19)(24)\,.
\label{eq:gluonmoment}
\end{align}
The numbers in the first parenthesis correspond to the statistical
error, the second is a systematic due to the excited states, and the third one
is systematic taken as the difference between the single- and double- smeared results; 
this is within the statistical errors.

Taking into account the disconnected quark contribution has small
effect on $\langle x \rangle^R_g$ due to the mild mixing when stout smearing
is applied on the gluon operator. Complete results on the quark and gluon
momentum fraction appear in Ref.~\cite{Alexandrou:2017oeh}.

\section{Conclusion and outlook}

In this paper we applied the direct method to compute the average
momentum fraction of the gluon in the nucleon, $\langle x \rangle_g$, 
taking into account the mixing with the singlet, light quark contribution. 
In order to obtain statistically significant results for the involved, 
purely disconnected 3-point functions,  several steps of stout smearing to the
gauge links that enter the operator were employed. Nevertheless, a substantial 
amount of measurements was needed to obtain a good signal with about 10\% 
statistical error. 

We computed the average momentum fraction for two gauge
field ensembles. The first has  $N_f{=}2{+}1{+}1$ flavors
representing the first two quark generations at a pion mass
of about 370\,MeV with 34470 measurements.  The second
ensemble has $N_f{=}2$  mass degenerate up and down
quarks at the physical value of the pion mass, with 204900 measurements.  The number of measurements
for the two cases allowed us to obtain statistically
significant values for the bare matrix elements (see
Eq.\,(\ref{eq_gm_res_bare})).

Since the required gluon operator is a singlet operator, it mixes
with the corresponding singlet quark operator. As a consequence, the renormalization 
of the gluon operator is highly non-trivial since this mixing has to be
taken into account. To this end, we have performed a perturbative 
calculation for the mixing and the renormalization. This has been 
done in the dimensional and the lattice regularizations. Moreover, the stout smearing that we
employed in the lattice computation of the bare matrix
element had to be taken into account in the perturbative 
calculation. This led to a very complicated perturbative calculation
which involved several  diagrams with $\mathcal O(100000)$ intermediate expressions. 
Still, we could demonstrate that with the inclusion of two stout 
smearing levels a saturation of the renormalization functions 
could be observed. The renormalization functions obtained in this manner have been 
used for the renormalization of gluon and the corresponding singlet 
quark moments. The final results for the renormalized gluon momentum
fraction are summarized in Eq.\,(\ref{eq:gluonmoment}), and in 
Ref.~\cite{Alexandrou:2017oeh} for the quark singlet quantities.
The values can serve for a comparison with a
phenomenological extraction of these quantities from deep
inelastic scattering experiments.  Our results also
demonstrate that the gluon indeed contributes a significant
amount of the momentum fraction of about 30\%. 

Our calculations can be extended to evaluate the spin
content of the nucleon, a topic we would like to report on
in the future. In addition, the renormalization functions
computed here can directly be used for the renormalization
of the corresponding average fractional momenta of the pion. 

\section*{Acknowledgments}
We thank our fellow members of ETMC for their constant collaboration. 
Helpful discussions with Fernanda Steffens, Keh-Fei Liu and 
Yi-Bo Yang are gratefully acknowledged.

We are grateful to the John von Neumann Institute for Computing (NIC),
the J{\"u}lich Supercomputing Center and the DESY Zeuthen Computing
Center for their computing resources and support. Computational resources
from the SwissNational Supercomputing Centre (CSCS) have also been used
under Projects No. s540 and s625.
This work has been supported in part by the Cyprus Research Promotion
Foundation through the  Project Cy-Tera
(Grant No. NEA Y$\Pi$O$\Delta$OMH/$\Sigma$TPATH/0308/31) 
co-financed by the European Regional Development Fund.

\pagebreak

\bibliographystyle{utphys}
\bibliography{./references}

\providecommand{\href}[2]{#2}\begingroup\raggedright\begin{thebibliography}{10}

\bibitem{Alexandrou:2014yha}
C.~Alexandrou, ``{Nucleon structure from lattice QCD - recent achievements and
  perspectives},'' \href{http://dx.doi.org/10.1051/epjconf/20147301013}{{\em
  EPJ Web Conf.} {\bfseries 73} (2014) 01013},
\href{http://arxiv.org/abs/1404.5213}{{\ttfamily arXiv:1404.5213 [hep-lat]}}.
%%CITATION = ARXIV:1404.5213;%%.

\bibitem{Constantinou:2014tga}
M.~Constantinou, ``{Hadron Structure},'' {\em PoS} {\bfseries LATTICE2014}
  (2014) 001,
\href{http://arxiv.org/abs/1411.0078}{{\ttfamily arXiv:1411.0078 [hep-lat]}}.
%%CITATION = ARXIV:1411.0078;%%.

\bibitem{Abdel-Rehim:2013wlz}
A.~Abdel-Rehim, C.~Alexandrou, M.~Constantinou, V.~Drach, K.~Hadjiyiannakou,
  K.~Jansen, G.~Koutsou, and A.~Vaquero, ``{Disconnected quark loop
  contributions to nucleon observables in lattice QCD},''
  \href{http://dx.doi.org/10.1103/PhysRevD.89.034501}{{\em Phys. Rev.}
  {\bfseries D89} no.~3, (2014) 034501},
\href{http://arxiv.org/abs/1310.6339}{{\ttfamily arXiv:1310.6339 [hep-lat]}}.
%%CITATION = ARXIV:1310.6339;%%.

\bibitem{Alexandrou:2013wca}
C.~Alexandrou, M.~Constantinou, V.~Drach, K.~Hadjiyiannakou, K.~Jansen, {\em
  et~al.}, ``{Evaluation of disconnected quark loops for hadron structure using
  GPUs},'' \href{http://dx.doi.org/10.1016/j.cpc.2014.01.009}{{\em
  Comput.Phys.Commun.} {\bfseries 185} (2014) 1370--1382},
\href{http://arxiv.org/abs/1309.2256}{{\ttfamily arXiv:1309.2256 [hep-lat]}}.
%%CITATION = ARXIV:1309.2256;%%.

\bibitem{Abdel-Rehim:2015owa}
A.~Abdel-Rehim {\em et~al.}, ``{Nucleon and pion structure with lattice QCD
  simulations at physical value of the pion mass},''
  \href{http://dx.doi.org/10.1103/PhysRevD.92.114513,
  10.1103/PhysRevD.93.039904}{{\em Phys. Rev.} {\bfseries D92} no.~11, (2015)
  114513}, \href{http://arxiv.org/abs/1507.04936}{{\ttfamily arXiv:1507.04936
  [hep-lat]}}.
[Erratum: {\it Phys. Rev.} {\bf D93} no. 3, (2016) 039904].
%%CITATION = ARXIV:1507.04936;%%.

\bibitem{Xiong:2013bka}
X.~Xiong, X.~Ji, J.-H. Zhang, and Y.~Zhao, ``{One-loop matching for parton
  distributions: Nonsinglet case},''
  \href{http://dx.doi.org/10.1103/PhysRevD.90.014051}{{\em Phys. Rev.}
  {\bfseries D90} no.~1, (2014) 014051},
\href{http://arxiv.org/abs/1310.7471}{{\ttfamily arXiv:1310.7471 [hep-ph]}}.
%%CITATION = ARXIV:1310.7471;%%.

\bibitem{Lin:2014zya}
H.-W. Lin, J.-W. Chen, S.~D. Cohen, and X.~Ji, ``{Flavor Structure of the
  Nucleon Sea from Lattice QCD},''
  \href{http://dx.doi.org/10.1103/PhysRevD.91.054510}{{\em Phys. Rev.}
  {\bfseries D91} (2015) 054510},
\href{http://arxiv.org/abs/1402.1462}{{\ttfamily arXiv:1402.1462 [hep-ph]}}.
%%CITATION = ARXIV:1402.1462;%%.

\bibitem{Alexandrou:2015rja}
C.~Alexandrou, K.~Cichy, V.~Drach, E.~Garcia-Ramos, K.~Hadjiyiannakou,
  K.~Jansen, F.~Steffens, and C.~Wiese, ``{Lattice calculation of parton
  distributions},'' \href{http://dx.doi.org/10.1103/PhysRevD.92.014502}{{\em
  Phys. Rev.} {\bfseries D92} no.~1, (2015) 014502},
\href{http://arxiv.org/abs/1504.07455}{{\ttfamily arXiv:1504.07455 [hep-lat]}}.
%%CITATION = ARXIV:1504.07455;%%.

\bibitem{Yang:2016plb}
Y.-B. Yang, R.~S. Sufian, A.~Alexandru, T.~Draper, M.~J. Glatzmaier, K.-F. Liu,
  and Y.~Zhao, ``{Glue Spin and Helicity in the Proton from Lattice QCD},''
  {\em Phys. Rev. Lett.} {\bfseries 118} no.~10, (2017) 102001,
\href{http://arxiv.org/abs/1609.05937}{{\ttfamily arXiv:1609.05937 [hep-ph]}}.
%%CITATION = ARXIV:1609.05937;%%.

\bibitem{Gockeler:1996zg}
M.~Gockeler, R.~Horsley, E.-M. Ilgenfritz, H.~Oelrich, H.~Perlt, P.~E.~L.
  Rakow, G.~Schierholz, A.~Schiller, and P.~Stephenson, ``{A Preliminary
  lattice study of the glue in the nucleon},''
  \href{http://dx.doi.org/10.1016/S0920-5632(96)00650-0}{{\em Nucl. Phys. Proc.
  Suppl.} {\bfseries 53} (1997) 324--326},
\href{http://arxiv.org/abs/hep-lat/9608017}{{\ttfamily arXiv:hep-lat/9608017
  [hep-lat]}}.
%%CITATION = HEP-LAT/9608017;%%.

\bibitem{Horsley:2012pz}
{\bfseries QCDSF, UKQCD} Collaboration, R.~Horsley {\em et~al.}, ``{A Lattice
  Study of the Glue in the Nucleon},''
  \href{http://dx.doi.org/10.1016/j.physletb.2012.07.004}{{\em Phys. Lett.}
  {\bfseries B714} (2012) 312--316},
\href{http://arxiv.org/abs/1205.6410}{{\ttfamily arXiv:1205.6410 [hep-lat]}}.
%%CITATION = ARXIV:1205.6410;%%.

\bibitem{Liu:2012nz}
K.~F. Liu {\em et~al.}, ``{Quark and Glue Momenta and Angular Momenta in the
  Proton --- a Lattice Calculation},'' {\em PoS} {\bfseries LATTICE2011} (2011)
  164,
\href{http://arxiv.org/abs/1203.6388}{{\ttfamily arXiv:1203.6388 [hep-ph]}}.
%%CITATION = ARXIV:1203.6388;%%.

\bibitem{Deka:2013zha}
M.~Deka {\em et~al.}, ``{Lattice study of quark and glue momenta and angular
  momenta in the nucleon},''
  \href{http://dx.doi.org/10.1103/PhysRevD.91.014505}{{\em Phys. Rev.}
  {\bfseries D91} no.~1, (2015) 014505},
\href{http://arxiv.org/abs/1312.4816}{{\ttfamily arXiv:1312.4816 [hep-lat]}}.
%%CITATION = ARXIV:1312.4816;%%.

\bibitem{Alekhin:2013nda}
S.~Alekhin, J.~Blumlein, and S.~Moch, ``{The ABM parton distributions tuned to
  LHC data},'' \href{http://dx.doi.org/10.1103/PhysRevD.89.054028}{{\em Phys.
  Rev.} {\bfseries D89} no.~5, (2014) 054028},
\href{http://arxiv.org/abs/1310.3059}{{\ttfamily arXiv:1310.3059 [hep-ph]}}.
%%CITATION = ARXIV:1310.3059;%%.

\bibitem{Frezzotti:2003ni}
R.~Frezzotti and G.~Rossi, ``{Chirally improving Wilson fermions. 1. O(a)
  improvement},'' \href{http://dx.doi.org/10.1088/1126-6708/2004/08/007}{{\em
  JHEP} {\bfseries 0408} (2004) 007},
\href{http://arxiv.org/abs/hep-lat/0306014}{{\ttfamily arXiv:hep-lat/0306014
  [hep-lat]}}.
%%CITATION = HEP-LAT/0306014;%%.

\bibitem{Frezzotti:2004wz}
R.~Frezzotti and G.~C. Rossi, ``{Chirally improving Wilson fermions. II.
  Four-quark operators},''
  \href{http://dx.doi.org/10.1088/1126-6708/2004/10/070}{{\em JHEP} {\bfseries
  10} (2004) 070},
\href{http://arxiv.org/abs/hep-lat/0407002}{{\ttfamily arXiv:hep-lat/0407002
  [hep-lat]}}.
%%CITATION = HEP-LAT/0407002;%%.

\bibitem{Meyer:2007tm}
H.~B. Meyer and J.~W. Negele, ``{Gluon contributions to the pion mass and light
  cone momentum fraction},''
  \href{http://dx.doi.org/10.1103/PhysRevD.77.037501}{{\em Phys. Rev.}
  {\bfseries D77} (2008) 037501},
\href{http://arxiv.org/abs/0707.3225}{{\ttfamily arXiv:0707.3225 [hep-lat]}}.
%%CITATION = ARXIV:0707.3225;%%.

\bibitem{Alexandrou:2013tfa}
C.~Alexandrou, V.~Drach, K.~Hadjiyiannakou, K.~Jansen, B.~Kostrzewa, and
  C.~Wiese, ``{Looking at the gluon moment of the nucleon with dynamical
  twisted mass fermions},'' {\em PoS} {\bfseries LATTICE2013} (2014) 289,
\href{http://arxiv.org/abs/1311.3174}{{\ttfamily arXiv:1311.3174 [hep-lat]}}.
%%CITATION = ARXIV:1311.3174;%%.

\bibitem{Ji:1994av}
X.-D. Ji, ``{A QCD analysis of the mass structure of the nucleon},''
  \href{http://dx.doi.org/10.1103/PhysRevLett.74.1071}{{\em Phys. Rev. Lett.}
  {\bfseries 74} (1995) 1071--1074},
\href{http://arxiv.org/abs/hep-ph/9410274}{{\ttfamily arXiv:hep-ph/9410274
  [hep-ph]}}.
%%CITATION = HEP-PH/9410274;%%.

\bibitem{Best:1997qp}
C.~Best, M.~Gockeler, R.~Horsley, E.-M. Ilgenfritz, H.~Perlt, P.~E.~L. Rakow,
  A.~Schafer, G.~Schierholz, A.~Schiller, and S.~Schramm, ``{Pion and rho
  structure functions from lattice QCD},''
  \href{http://dx.doi.org/10.1103/PhysRevD.56.2743}{{\em Phys. Rev.} {\bfseries
  D56} (1997) 2743--2754},
\href{http://arxiv.org/abs/hep-lat/9703014}{{\ttfamily arXiv:hep-lat/9703014
  [hep-lat]}}.
%%CITATION = HEP-LAT/9703014;%%.

\bibitem{Guagnelli:2004ga}
{\bfseries Zeuthen-Rome (ZeRo)} Collaboration, M.~Guagnelli, K.~Jansen,
  F.~Palombi, R.~Petronzio, A.~Shindler, and I.~Wetzorke, ``{Non-perturbative
  pion matrix element of a twist-2 operator from the lattice},''
  \href{http://dx.doi.org/10.1140/epjc/s2005-02121-5}{{\em Eur. Phys. J.}
  {\bfseries C40} (2005) 69--80},
\href{http://arxiv.org/abs/hep-lat/0405027}{{\ttfamily arXiv:hep-lat/0405027
  [hep-lat]}}.
%%CITATION = HEP-LAT/0405027;%%.

\bibitem{Baron:2010bv}
R.~Baron {\em et~al.}, ``{Light hadrons from lattice QCD with light (u,d),
  strange and charm dynamical quarks},''
  \href{http://dx.doi.org/10.1007/JHEP06(2010)111}{{\em JHEP} {\bfseries 1006}
  (2010) 111},
\href{http://arxiv.org/abs/1004.5284}{{\ttfamily arXiv:1004.5284 [hep-lat]}}.
%%CITATION = ARXIV:1004.5284;%%.

\bibitem{Alexandrou:2014sha}
C.~Alexandrou, V.~Drach, K.~Jansen, C.~Kallidonis, and G.~Koutsou, ``{Baryon
  spectrum with $N_f=2+1+1$ twisted mass fermions},''
  \href{http://dx.doi.org/10.1103/PhysRevD.90.074501}{{\em Phys. Rev.}
  {\bfseries D90} no.~7, (2014) 074501},
\href{http://arxiv.org/abs/1406.4310}{{\ttfamily arXiv:1406.4310 [hep-lat]}}.
%%CITATION = ARXIV:1406.4310;%%.

\bibitem{Abdel-Rehim:2015pwa}
{\bfseries ETM} Collaboration, A.~Abdel-Rehim {\em et~al.}, ``{First physics
  results at the physical pion mass from $N_f=2$ Wilson twisted mass fermions
  at maximal twist},'' {\em Phys. Rev.} {\bfseries D95} no.~9, (2017) 094515,
\href{http://arxiv.org/abs/1507.05068}{{\ttfamily arXiv:1507.05068 [hep-lat]}}.
%%CITATION = ARXIV:1507.05068;%%.

\bibitem{Alexandrou:2013joa}
C.~Alexandrou, M.~Constantinou, S.~Dinter, V.~Drach, K.~Jansen, {\em et~al.},
  ``{Nucleon form factors and moments of generalized parton distributions using
  $N_f=2+1+1$ twisted mass fermions},''
  \href{http://dx.doi.org/10.1103/PhysRevD.88.014509}{{\em Phys. Rev.}
  {\bfseries D88} no.~1, (2013) 014509},
\href{http://arxiv.org/abs/1303.5979}{{\ttfamily arXiv:1303.5979 [hep-lat]}}.
%%CITATION = ARXIV:1303.5979;%%.

\bibitem{Hasenfratz:2001hp}
A.~Hasenfratz and F.~Knechtli, ``{Flavor symmetry and the static potential with
  hypercubic blocking},''
  \href{http://dx.doi.org/10.1103/PhysRevD.64.034504}{{\em Phys. Rev.}
  {\bfseries D64} (2001) 034504},
\href{http://arxiv.org/abs/hep-lat/0103029}{{\ttfamily arXiv:hep-lat/0103029
  [hep-lat]}}.
%%CITATION = HEP-LAT/0103029;%%.

\bibitem{Morningstar:2003gk}
C.~Morningstar and M.~J. Peardon, ``{Analytic smearing of SU(3) link variables
  in lattice QCD},'' \href{http://dx.doi.org/10.1103/PhysRevD.69.054501}{{\em
  Phys. Rev.} {\bfseries D69} (2004) 054501},
\href{http://arxiv.org/abs/hep-lat/0311018}{{\ttfamily arXiv:hep-lat/0311018
  [hep-lat]}}.
%%CITATION = HEP-LAT/0311018;%%.

\bibitem{Joglekar:1976}
S.~Joglekar and B.~Lee, ``{General Theory of Renormalization of Gauge Invariant
  Operators},'' {\em Ann.Phys.} {\bfseries 97} (1976) 160.

\bibitem{Alexandrou:2010me}
C.~Alexandrou, M.~Constantinou, T.~Korzec, H.~Panagopoulos, and F.~Stylianou,
  ``{Renormalization constants for 2-twist operators in twisted mass QCD},''
  \href{http://dx.doi.org/10.1103/PhysRevD.83.014503}{{\em Phys. Rev.}
  {\bfseries D83} (2011) 014503},
\href{http://arxiv.org/abs/1006.1920}{{\ttfamily arXiv:1006.1920 [hep-lat]}}.
%%CITATION = ARXIV:1006.1920;%%.

\bibitem{GluonPertRenorm}
M.~Constantinou and H.~Panagopoulos. In preparation.

\bibitem{Caracciolo:1991cp}
S.~Caracciolo, P.~Menotti, and A.~Pelissetto, ``{One loop analytic computation
  of the energy momentum tensor for lattice gauge theories},''
\href{http://dx.doi.org/10.1016/0550-3213(92)90339-D}{{\em Nucl. Phys.}
  {\bfseries B375} (1992) 195--239}.
%%CITATION = NUPHA,B375,195;%%.

\bibitem{Constantinou:2015ela}
M.~Constantinou, M.~Costa, R.~Frezzotti, V.~Lubicz, G.~Martinelli, D.~Meloni,
  H.~Panagopoulos, and S.~Simula, ``{Renormalization of the chromomagnetic
  operator on the lattice},''
  \href{http://dx.doi.org/10.1103/PhysRevD.92.034505}{{\em Phys. Rev.}
  {\bfseries D92} no.~3, (2015) 034505},
\href{http://arxiv.org/abs/1506.00361}{{\ttfamily arXiv:1506.00361 [hep-lat]}}.
%%CITATION = ARXIV:1506.00361;%%.

\bibitem{Alexandrou:2016tuo}
C.~Alexandrou, M.~Constantinou, K.~Hadjiyiannakou, C.~Kallidonis, G.~Koutsou,
  K.~Jansen, C.~Wiese, and A.~V. Aviles-Casco, ``{Nucleon spin and quark
  content at the physical point},'' {\em PoS} {\bfseries LATTICE2016} (2016)
  153,
\href{http://arxiv.org/abs/1611.09163}{{\ttfamily arXiv:1611.09163 [hep-lat]}}.
%%CITATION = ARXIV:1611.09163;%%.

\bibitem{Abdel-Rehim:2016pjw}
A.~Abdel-Rehim, C.~Alexandrou, M.~Constantinou, J.~Finkenrath,
  K.~Hadjiyiannakou, K.~Jansen, C.~Kallidonis, G.~Koutsou, A.~V. Aviles-Casco,
  and J.~Volmer, ``{Disconnected diagrams with twisted-mass fermions},'' {\em
  PoS} {\bfseries LATTICE2016} (2016) 155,
\href{http://arxiv.org/abs/1611.03802}{{\ttfamily arXiv:1611.03802 [hep-lat]}}.
%%CITATION = ARXIV:1611.03802;%%.

\bibitem{Alexandrou:2017oeh}
C.~Alexandrou, M.~Constantinou, K.~Hadjiyiannakou, K.~Jansen, C.~Kallidonis,
  G.~Koutsou, A.~V. Aviles-Casco, and C.~Wiese, ``{Nucleon spin and momentum
  decomposition using lattice QCD simulations},'' {\em accepted in Phys. Rev.
  Lett.} (2017) ,
\href{http://arxiv.org/abs/1706.02973}{{\ttfamily arXiv:1706.02973 [hep-lat]}}.
%%CITATION = ARXIV:1706.02973;%%.

\end{thebibliography}\endgroup

\end{document}